\documentclass[letterpaper,5p,times]{elsarticle}\usepackage{etoolbox}\newtoggle{preprint}\toggletrue{preprint} 

\usepackage[utf8]{inputenc}
\usepackage{textcomp}
\usepackage{amsmath}
\usepackage{siunitx}[2023-02-21 v3.2.1]\DeclareSIUnit\KWH{kWh}\DeclareSIUnit\year{yr}\iftoggle{preprint}{\providecommand\qty{\SI}\providecommand\qtyrange{\SIrange}}{\sisetup{text-series-to-math=true,reset-math-version = false}} 
\usepackage{pgfplots,tikz}\pgfplotsset{compat=1.14}
\usetikzlibrary{math}
\usetikzlibrary{datavisualization, datavisualization.formats.functions}
\usepackage{footnotebackref}

\usepackage{cleveref}
\usepackage{xintexpr}\xintverbosetrue\xintDigits:=16\relax 
\newcommand\FVof{}
\def\FVof(#1){\pgfmathfloatvalueof{#1}}

\iftoggle{preprint}{\journal{arXiv}}{}

\DeclareMathOperator{\erfc}{erfc}

\newcommand\kpWavelength{\qty{1.06}{\micro\meter}}
\newcommand\kSpeed{\qty{.2}{c}}
\newcommand\kpEmittance{0.01}
\newcommand\kpAbsorptance{$10^{-8}$}
\newcommand\kpMaxTemp{\qty{625}{\kelvin}}

\newcommand\kpMapsPayloadMin{\qty{0.1}{\milli\gram}}
\newcommand\kpMapsPayloadMax{\qty{100}{\kilo\tonne}}
\newcommand\kpMapsCruiseRangeC{\qtyrange{.0001}{.99}{c}}
\newcommand\kpMapsCruiseRangeAuYr{\qtyrange{6.3}{63000}{au\per\year}}
\newcommand\kpMapsLaserDiaMax{\qty{10}{\kilo\meter}}
\newcommand\kpMapsGridP{\qty{5}{\giga\watt}}

\newcommand\pAMass{\qty{10}{\kilo\gram}}
\newcommand\pASpeedc{\qty{.001}{c}}
\newcommand\pASpeedau{\qty{63}{au\per\year}}
\newcommand\pAGCapex{\$26B}
\newcommand\pAOCapex{\$610M}
\newcommand\pAOSIDia{\qty{1.2}{\kilo\meter}}
\newcommand\pAOSIIDia{\qty{77}{\meter}}
\newcommand\pAOPower{\qty{2.5}{\giga\watt}}
\newcommand\pAOOpex{\$58M}
\newcommand\pAOAccnInit{\qty{0.08}{g's}}

\newcommand\pBMass{\qty{10}{\kilo\gram}}
\newcommand\pBSpeedc{\qty{.0001}{c}}
\newcommand\pBSpeedau{\qty{6.3}{au\per\year}}
\newcommand\pBGCapex{\$2.6B}
\newcommand\pBOCapex{\$17M}
\newcommand\pBOOpex{\$5.7M}

\newcommand\pCMass{\qty{0.1}{\milli\gram}}
\newcommand\pCSpeedc{\qty{.01}{c}}
\newcommand\pCGCapex{\$25M}
\newcommand\pCOCapex{\$21M}
\newcommand\pCOSIIDia{\qty{17}{\centi\meter}}
\newcommand\pCOOpex{\$12k}

\newcommand\pDMass{\qty{100}{\kilo\tonne}}
\newcommand\pDSpeedc{\qty{0.07}{c}}
\newcommand\pDSpeedau{\qty{4400}{au\per\year}}
\newcommand\pDGCapex{\$190B/kg}
\newcommand\pDOCapex{\$190B/kg}
\newcommand\pDGOpex{\$370M/kg}
\newcommand\pDOOpex{\$370M/kg}
\newcommand\pDOSIIDia{\qty{7.4}{\kilo\meter}}
\newcommand\pDOPower{\qty{380}{\peta\watt}}
\newcommand\pDODurAccn{\qty{21}{\day}}
\newcommand\pDODurPul{\qty{20}{\day}}

\begin{document} 

\iftoggle{preprint}{\begin{frontmatter}}{}
\title{Cost-Optimal Laser-Accelerated Lightsails}
\iftoggle{preprint}{
\author{Kevin L. G. Parkin\corref{kpfn}}\ead{kevin@parkinresearch.com}\cortext[kpfn]{Systems Director, Breakthrough Starshot}
}{\author{Kevin L. G. Parkin\footnote{Systems Director, Breakthrough Starshot}}\affil{Parkin Research LLC, 2261 Market Street \#221, San Francisco, USA. kevin@parkinresearch.com}}

\iftoggle{preprint}{}{\maketitle}

\begin{abstract}

Laser-accelerated lightsails enable new types of missions that are very different from the Breakthrough Starshot mission to the Centauri system that aims to send \qty{1}{\gram} of payload at \kSpeed. The present work widens the mission design space to \kpMapsPayloadMin\ to \kpMapsPayloadMax\ payload and \kpMapsCruiseRangeC\ cruise velocity. Drawing up to \kpMapsGridP\ directly from the grid (to augment power drawn from local energy storage) turns out to be the key to making small missions affordable: It collapses the accelerating laser's capital cost by up to 5 orders of magnitude, enabling new possibilities such as a \pAMass\ Solar system cubesat that accelerates to \pASpeedc\ (\pASpeedau) using a \pAOSIIDia\ sail and \pAOCapex\ laser, costing \pAOOpex\ worth of energy per mission.

Trajectory equations describing lightsail acceleration are derived in closed form and used instead of numerical integration. Consequently, analyses have progressed from single point designs to whole performance maps comprised of thousands of cost-optimized point designs. The performance maps reveal qualitatively different regimes characterized by the particular constraint that drives cost, and these driving constraints change depending on mission payload mass and cruise velocity.

The performance maps also reveal a family of cost-optimal missions that accelerate at Earth gravity: The heaviest such mission is a \pDOSIIDia\ diameter \pDMass\ vessel (equivalent to 225 International Space Stations) that is accelerated for \pDODurAccn\ to achieve \pDSpeedc, reaching the Centauri system within a human lifetime. While unthinkable at this time, the required \pDOPower\ peak radiated power (twice terrestrial insolation) might be generated by space solar power or fusion within a few centuries. Regardless, it is now possible to contemplate such missions using laser-accelerated lightsails.

\iftoggle{preprint}{}{}

\end{abstract}\iftoggle{preprint}{\end{frontmatter}}{}

\section{Motivation}

\iftoggle{preprint}{A}{\lettrine{A}} system-level model was developed to compute a cost-optimal design for a laser-accelerated lightsail mission to carry \qty{1}{\gram} of payload to the Centauri system at \kSpeed\ for the Breakthrough Starshot initiative. The model and mission design were published in 2018 \cite{parkin2018breakthrough}. Since then, there has been a growing need to compute point designs for heavier missions at other cruise velocities.

In 2020, Witten \cite{witten2020searching} suggested a phalanx of low-cost laser-launched lightsails to probe for the gravitational field of a primordial black hole in the outer Solar system; an idea that was also examined by Christian, Hoang, and Loeb \cite{christian2017interferometric,Hoang_2020}. The lightsails would be like the ones envisioned for Starshot, but cruise more slowly at \qty{.001}{c} and carry orders of magnitude heavier payloads. Hoang \& Loeb \cite{Hoang_2020} estimated that at this cruise velocity, a spacecraft should weigh more than \qty{10}{\kilo\gram} for the gravitational signal to be distinguishable from drag noise and magnetic noise. In response, the system model was used to generate point designs for \qty{.001}{c} (\qty{63}{au\per\year}) precursor missions carrying \qty{1}{\kilo\gram}, \pAMass, and \qty{100}{\kilo\gram} payloads \cite{turyshev2020exploration}. 

In FY2017, the United States Congress directed NASA to undertake an interstellar mission technology assessment \cite{culbertson2016} toward the goal of launching a probe to Alpha Centauri by the year 2069 in commemoration of the one-hundredth anniversary of the Apollo-11 moon landing. NASA responded by holding three extramural workshops to identify and evaluate technology concepts. Litchford \& Sheehy \cite{litchford2020prospects} summarized these activities and suggested phased science goals and missions to incrementally develop the interstellar propulsion, spacecraft subsystems, and operational experience, needed to mount a fully-integrated scientific probe to nearby stars. Knowing how laser-driven lightsails scale to heavier and slower missions spanning the inner Solar system through to the Oort cloud would benefit planning of this kind.

Breakthrough Starshot has also needed to compute heavier and slower mission point designs: Ever since the initiative launched in 2016, interstellar precursor mission concepts have been collected and viewed as potential stepping stones toward the main Centauri system mission. For example, an early need is to send sub-scale lightsails to probe the nearby interstellar magnetic field and dust environment ahead of the main mission. Another precursor mission candidate has been the solar gravitational lens mission \cite{turyshev2020direct}, whose \qty{550}{au} starting distance can be reached by laser-driven lightsails within a decade. This mission's need for precision maneuvering and imaging sensors is shared by a class of missions that catch up with interstellar objects passing through the Solar system \cite{hein2022interstellar}. 

These and other mission concepts involve slower cruise velocities and very different payload masses than the Centauri system mission, making them difficult to compute using the earlier Starshot system model: The model's numerical solvers relied on manually-adjusted search ranges that needed re-tuning for different mission parameters. Due to computational complexity, the model converged only very slowly upon each point design, sometimes taking an hour or more before hitting a variable's limit, which then misguided the cost optimizer. In that case, the manual re-tuning process would begin again. A faster system model was needed, with improved convergence and autonomy over a wider range of mission parameters. This is what motivated the present work.

\section{Changes Since the 2018 System Model}

At its core, the system model describes the propagation of a beam from a ground-based laser to a spaceborne lightsail, and the lightsail's resulting motion. It relates key design parameters that determine the system capex (capital expense) and opex (operational expense). Optimizers automatically vary the size and power of the laser and the size of the lightsail to minimize opex and capex. 

Compared to the 2018 system model \cite{parkin2018breakthrough}, aspects are changed as follows.

\subsection{Terminology}\label{sec:Terminology}

`Laser' denotes the ground-based integrated facility comprised of laser sources, optics, optoelectronics, energy storage and cooling subsystems, whose emissions coherently combine at the intended focus. Elsewhere this is called the `photon engine', the `beamer' or `beam director', all of which are mildly problematic. This paper uses `laser source' to refer to just the part that converts electrical power to laser power, without the supporting optics, energy storage, cooling or other subsystems. 

`Lightsail' denotes the integrated sail material, spacecraft and payload. Formerly this was called the sailcraft (sail and craft, which may be physically discrete or integrated). This paper uses `sail' to refer to just the sail part used for propulsion by reflecting incident light, without the spacecraft functionality or payload. 

\subsection{Nomenclature}\label{sec:Nomenclature}

\begin{figure}[h]\centering\includegraphics[width=\columnwidth]{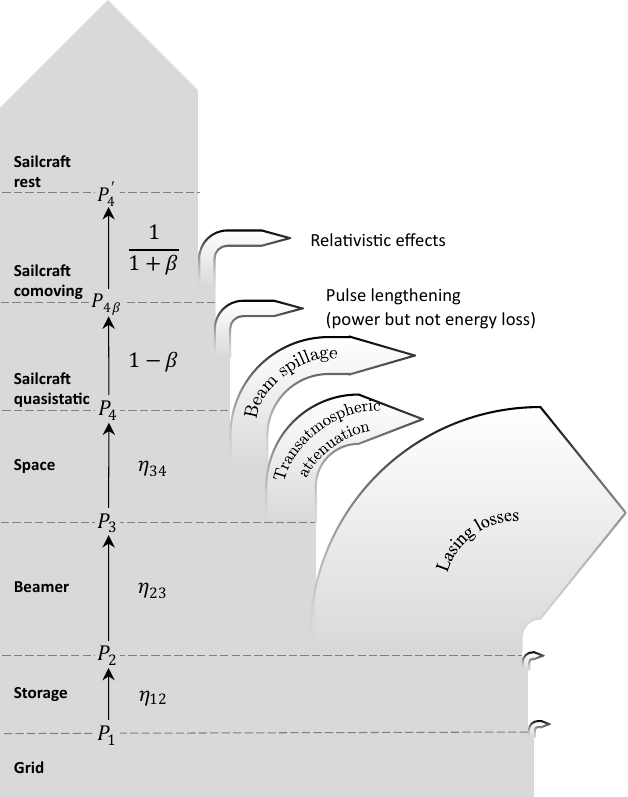}\caption{Power and efficiency relationships}\label{fig:PowerEfficiency}\end{figure}

Powers and efficiencies are renamed with subscripts $1$ to $4$ corresponding to stages of passage from electrical grid through to lightsail. This sequence is shown in \cref{fig:PowerEfficiency}, a Sankey diagram that is extended to include frame transformations.

Following the figure from grid through to lightsail, a grid supplies power $P_{1}$ to an energy storage facility, corresponding to `wallplug' power at the organizational interface between the provider and Starshot as its customer. After minor losses due energy conversion, storage, and local transmission, the energy storage subsystem supplies power $P_2$ to the laser. Within the laser, each laser source's electrical efficiency is relatively low, and there are various smaller loss mechanisms. After losses, the laser radiates power in the direction of a lightsail. Radiated power $P_{3}$ relates to other powers via,\begin{equation}P_{3}=\eta _{23}P_{2}=\eta _{12}\eta _{23}P_{1}.\label{eq:P3P3P1}\end{equation}

Power transfer efficiencies $\eta _{12}$ and $\eta _{23}$ also relate energies (in addition to powers) from the grid to the laser because they are invariant over the course of a pulse. Through space and thereafter, the efficiencies vary with time.

Between the laser and the lightsail there is a delay due to the finite speed of light, given by \begin{equation}\label{eq:P3P4}\frac{P_{4}(t)}{\eta _{34}(t)}=P_{3}\left( t-\frac{z(t)}{c}\right).\end{equation} The delay $z(t)/c$ lengthens as the lightsail accelerates away. If $t$=0 when the first photons reach the lightsail, then the photons' path losses (1-$\eta_{34}$) depend on the lightsail's initial separation from the laser $z_{0}\equiv z(t$=$0)$. This means that the first photons depart the laser at time $-z_{0}/c$.

Power reaching the lightsail can be viewed in different ways, each of which has particular utility: $P_4$, the power received in the quasistatic frame, is used with power transfer efficiency calculations that assume a static target relative to the source, such as the Goubau equations in the next section. $P_{4\beta}$, the power received in the lightsail co-moving frame, is integrated with respect to time to find the radiated energy $Q_3$ (see \cref{section:RadiatedEqns}). $P_{4}^{\prime}$, the power received in the lightsail relativistic rest frame, can be constant when the lightsail is accelerating at its thermal limit (see \cref{section:TLimitedEqns}). These lightsail powers are related by \begin{equation}\label{eq:P4bP4}P_{4\beta}=(1-\beta)P_{4},\end{equation}\begin{equation}\label{eq:P4pP4b}P_{4}^{\prime }=\frac{1}{1+\beta }P_{4\beta }.\end{equation} where factor $1$-$\beta$ is the fractionally-reduced rate at which photons hit the receding lightsail, and factor $1/(1+\beta)$ is the Doppler frequency downshift of laser-radiated photons as perceived by the receding lightsail. This latter factor is also the fractional photon energy loss, because photon frequency $\nu$ is linearly proportional to photon energy ($E_{\nu}$=$h\nu$). Multiplying the photon energy factor by the rate factor yields the power factor derived by Einstein \cite{einstein1905elektrodynamik} relating power between frames, \begin{equation}P_{4}^{\prime }=\frac{1-\beta }{1+\beta }P_{4}.\label{eq:einsteinpower0}\end{equation}

\subsection{Goubau Beam} \label{sec:goubau}

\xintdeffloatfunc xatau(t):=sqrt(twoPi/t);
\xintdeffloatfunc xbtau(t):=exp(twoPi/t);
\xintdeffloatfunc xetao(a,b):=((a^4+sqrt(a^8-4*b*a^4+4*b^2-8*b+4))/2/b)^2;
\xintdeffloatfunc xetat(a):=(0.5*a^2-(a^6)/32+(a^10)/4608*7)^2; 
\xintdeffloatfunc xetas(t):=if(t==0,1.0,(xatau(t)>1.2174805119418)*xetao(xatau(t),xbtau(t))+(t!=0)*(xatau(t)<=1.2174805119418)*xetat(xatau(t)));
\xintdeffloatfunc xtaud(t):=t; 
\xintdeffloatfunc xetad(t):=if(t<1.0E-5,0.0,1.0-exp(-t^2)); 
\xintdeffloatfunc xetak(t):=if(t<1.0E-5,0.0,xetas(Pi/t));
\xintdeffloatfunc xetamag(t):=(xetad(t)-xetak(t))/xetak(t);

\begin{figure}[h]
\begin{tikzpicture}[ 
declare function={
	etas(\t)    = {\xintfloateval{xetak(\FVof(\t))}}; 
	etad(\t)    = {\xintfloateval{xetad(\FVof(\t))}}; 
}]
\begin{axis}[
	axis x line=bottom,
	axis y line=left,
	grid=both,
    x label style={at={(axis description cs:1.05,-0.01)}, anchor=center},
	xlabel={$\tau_{d}$},
	ylabel={$\eta_b\left(\tau\right)$},
	domain=0.01:5,ymin=0,ymax=1.1,xmin=0,xmax=5.5,
	]
\addplot[color=red,samples=500]{etad(x)};
\addplot[color=blue,samples=500]{etas(x)};	
\end{axis}
\begin{axis}[
	axis x line shift = +0.15, 
	x label style={at={(axis description cs:1.05,-0.14)}, anchor=center},
	axis lines=left,
    xtick={0,1,2,3,4,5},
    xticklabels={$\infty$,$\pi$,$\frac{\pi}{2}$,$\frac{\pi}{3}$,$\frac{\pi}{4}$,$\frac{\pi}{5}$},
	xlabel = {$\tau_{g}$},
	ylabel=,
	ytick=,
	yticklabels=,
	domain=0.01:5,ymin=0,ymax=1.1,xmin=0,xmax=5.5,
	]
\end{axis}	
\end{tikzpicture}
\begin{tikzpicture}[ 
declare function={
	etamag(\t)  = {\xintfloateval{xetamag(\FVof(\t))}};
}]
\begin{axis}[
	axis x line=bottom,
	axis y line=left,
	grid=both,
    x label style={at={(axis description cs:1.05,-0.01)}, anchor=center},
	xlabel = {$\tau_{d}$},
	ylabel={$(\eta_{d}-\eta_{p})/\eta_{p}$},
	domain=0.01:5,ymin=-0.015,ymax=0.03,xmin=0,xmax=5.5,
	] %
\addplot[color=blue,samples=500]{etamag(x)};
\end{axis}
\begin{axis}[
	axis x line shift = +0.15, 
	x label style={at={(axis description cs:1.05,-0.14)}, anchor=center},
	axis lines=left,
    xtick={0,1,2,3,4,5},
    xticklabels={$\infty$,$\pi$,$\frac{\pi}{2}$,$\frac{\pi}{3}$,$\frac{\pi}{4}$,$\frac{\pi}{5}$},
	xlabel = {$\tau_{g}$},
	ylabel=,
	ytick=,
	yticklabels=,
	domain=0.01:5,ymin=0,ymax=1.1,xmin=0,xmax=5.5,
	]
\end{axis}	
\end{tikzpicture}
\caption{Top: Goubau beam power transfer efficiency calculated using Dickinson's approximation, \cref{eq:dickinsonefficiency}, in red, overplotted by \cref{eq:goubauefficiency}, in blue. Bottom: Fractional difference in power transfer efficiency approximations.}\label{fig:goubaudiff}\end{figure}
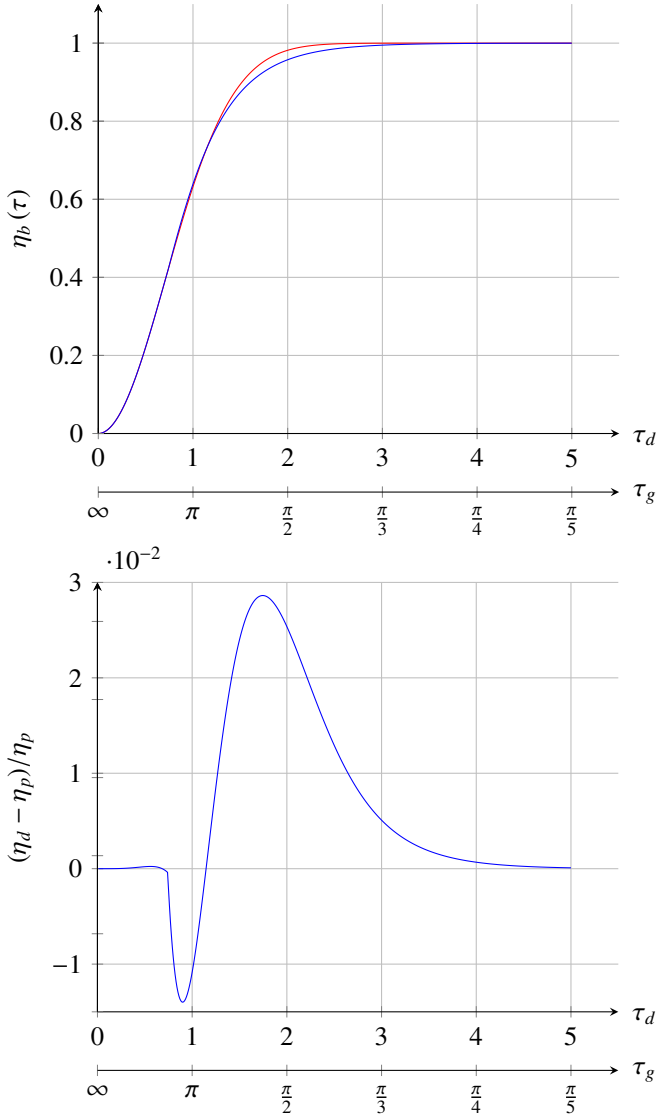

A Goubau beam \cite{goubau19683} describes near-optimal power transfer between \textit{finite diameter} optics \cite{hansen2005universal}. It is appropriate when the intensity distribution across a transmitting optic can be tailored. This is the case for a phased array, which consists of very many elements whose intensities are individually controlled and varied for optimal performance throughout the lightsail's acceleration. A Gaussian beam is a poor fit for this situation because it is accurate only in the limit of energy transfer between \textit{infinite diameter} optics, resulting in oversized optics costing billions of dollars extra. A top-hat beam describes energy transfer between a finite diameter optic and an infinite diameter optic and is also non-optimal when optics' capex dominates, as it can do for laser-driven lightsail.

Goubau beam transfer efficiency $\eta_{g}$ can be approximated by\begin{equation}
\eta_{g} (a)=\left\lbrace 
\begin{array}{ll}
\eta _{g+}(a) & \text{if }a>1.21748051194181 \\ 
\eta _{g-}(a) & \text{otherwise}
\end{array} 
\right.
\label{eq:goubauefficiency}\end{equation}
\begin{equation}\eta _{g+}(a) \equiv \frac{1}{4b^2}\left( a^{4}+\allowbreak \sqrt{a^{8}-4a^{4}b+4b^{2}-8b+4}\allowbreak \right)^{2}\end{equation}
\begin{equation}\eta _{g-}(a) \equiv \left( \frac{a^{2}}{2}-\frac{a^{6}}{32}+\frac{7a^{10}}{4608}\right) ^{2},\label{eq:goubaueta2}\end{equation}
where $a(\tau_{g}) \equiv \sqrt{\frac{2\pi }{\tau_{g}}}$ and $b\equiv e^{a^{2}}$. In this expression, transfer efficiency depends on just a single dimensionless parameter $\tau_{g}$, which in turn depends on the product of (circular) optic radii $R_{s}$ and $R_{b}$, their separation $z$, and beam wavelength $\lambda$. The expression for $\tau_{g}$ used in earlier work \cite{parkin2018breakthrough, parkin2017mtpfinalreport} contains an erroneous factor of two that traces back to the caption of Figure 2 in Goubau's book chapter \cite{goubau19683}. The correct expression for $\tau_{g}$ is\begin{equation}\tau_{g} \equiv \frac{\pi\lambda z}{\sqrt{\mathcal{A}_{3}\mathcal{A}_{4}}}=\frac{\lambda z}{R_{3}R_{4}}.\label{eq:goubautaudef}\end{equation}Earlier numerical results are unaffected because the code did not directly calculate $\tau_{g}$, instead calculating $a$ using an alternate (correct) expression.

Subsequent to Goubau \cite{goubau19683}, works by Brown \cite{brown1992beamed}; Hansen, McSpadden, and Benford \cite{hansen2005universal}; and Goubau himself \cite{goubau1970} use a different dimensionless grouping, also usually called $\tau$, to parameterize transfer efficiency, \begin{equation}\tau _{d}\equiv\frac{\sqrt{\mathcal{A}_{3}\mathcal{A}_{4}}}{\lambda z}=\frac{\pi }{\tau _{g}}=\frac{\pi R_{3}R_{4}}{\lambda z}=\frac{z_{GR}}{z}.\label{eq:dickinsontau}\end{equation} Here, the expression is reduced to a simple ratio by introducing a geometric mean of Rayleigh ranges $z_{GR}\equiv\pi R_{3}R_{4}/\lambda$: If the laser were to emit a Gaussian beam, its Rayleigh range would be $\pi R_{3}^{2}/\lambda$, and if the lightsail were to emit a Gaussian beam its Rayleigh range would be $\pi R_{4}^{2}/\lambda$, so $z_{GR}$ is the geometric mean of these. Subscripts have also been added to the two different versions of $\tau$ to clarify which is which and how they relate to each other.

For completeness, a dimensionless grouping\begin{equation}N_{T}\equiv R_{3}R_{4}/\lambda z=\tau_{d}/\pi\end{equation}is occasionally called the target (or sail) Fresnel number and used instead of $\tau_{d}$ or $\tau_{g}$. It is distinct from the conventional Fresnel number, which does not incorporate target size.

The system model now adopts the use of $\tau _{d}$ (instead of $\tau _{g}$ or $N_{T}$) because $\tau _{d}$ lends itself to a greatly simplified expression for transfer efficiency:\begin{equation}\eta _{d}=1-e^{-\tau_{d}^{2}}.\label{eq:dickinsonefficiency}\end{equation}
The lineage of this approximation is unclear \cite{dickinson2019originoftau, benford2022originoftau}; it is stated without proof by Benford starting in 2007 \cite{benford2007transferefficiency,benford2015transferefficiency}, but Dickinson and a colleague at NASA JPL were likely the first to deduce it (in 1998 or prior) by fitting it to the curve in Figure 2 of the 1970 article by Goubau \cite{goubau1970}. Regardless, \Cref{fig:goubaudiff} of the present paper shows that \cref{eq:dickinsonefficiency} differs from \cref{eq:goubauefficiency} by less than 3\%. \Cref{eq:dickinsonefficiency} is now used instead of \cref{eq:goubauefficiency} because it is simpler and quicker to calculate. 

Finally, the power transfer efficiency from the laser to the lightsail is given by,\begin{equation}\label{eq:eta34}\eta _{34}(t)=\eta _{a}\eta _{d}(z(t)),\end{equation}where $\eta _{a}$ accounts for scattering and absorption losses as the beam transits the atmosphere, and $\eta _{d}$ is calculated from \cref{eq:dickinsonefficiency,eq:dickinsontau} above.

\subsection{Closed-Form Trajectory Equations}

The quasi-1D motion\footnote{F{\H{u}}zfa et al. \cite{fHuzfa2020sailing} consider the more general case of a nonequilibrium lightsail in nonrectilinear motion.} of a lightsail in thermal equilibrium in its rest frame is described by \cite{parkin2018breakthrough},
\begin{equation}F=(A+2R) \frac{P_{4}^{\prime }}{c},\label{eq:eqnmotionrestframe}\end{equation}where $F$ is the force acting on the lightsail due to the incident beam power $P_{4}^{\prime }$, $A$ is lightsail absorptance, $R$ is lightsail reflectance, and $c$ is the speed of light in free space.

To relate $P_{4}^{\prime }$ to laser and beam propagation characteristics, the equation of motion is transformed from the lightsail rest frame to the quasistatic frame: Force remains unchanged because a pure force (that does not add net heat to the lightsail) is equal in both frames. In the quasistatic frame, force is related to acceleration by\begin{equation}F=\frac{dp}{dt}=\frac{d}{dt}(\gamma m_{0}\beta c) =m_{0}c\frac{d( \beta \gamma) }{dt}=\frac{E_{0}}{c}\gamma ^{3}\frac{d\beta }{dt},\label{eq:eqnmotionlaserframe}\end{equation}where $p$ is relativistic momentum, $t$ is time measured in the quasistatic frame, $\beta$ is lightsail speed as a fraction of light speed, $\gamma\equiv 1/\sqrt{1-\beta ^{2}}$ is the Lorentz factor, $m_{0}$ is the lightsail rest mass, and $E_{0}$ is the lightsail rest energy famously equal to $m_{0}c^{2}$. 

Equating \cref{eq:eqnmotionrestframe} to \cref{eq:eqnmotionlaserframe} yields the equation of motion in a more convenient form,\begin{equation}\frac{E_{0}}{A+2R}\frac{d\beta }{dt}=\frac{P_{4}^{\prime }}{\gamma ^{3}}=\frac{P_{4}}{D^{2}\gamma ^{3}},\label{eq:eqnmotionparkin}\end{equation}where the square of the Doppler factor $D^{2}$=$(1+\beta)/(1-\beta) $ converts $P_{4}^{\prime }$ to $P_{4}$ as per \cref{eq:einsteinpower0}.

To find the lightsail's distance from the laser $z(t)$, the earlier system model \cite{parkin2018breakthrough} uses the definition of $\beta$,\begin{equation}\beta\equiv\frac{1}{c}\frac{dz}{dt},\label{eq:eqnbeta}\end{equation}to transform \cref{eq:eqnmotionparkin} into a second order ODE,\begin{equation}\frac{d^{2}z}{dt^{2}}=\frac{c(A+2R)}{E_{0}}\frac{P_{4}}{D^{2}\gamma ^{3}},\end{equation}numerically integrating it to obtain $z(t)$ and auxiliary quantities.

In contrast, the present work uses the identity,\begin{equation}\frac{d\beta }{dt}=\frac{dz}{dt}\frac{d\beta }{dz}=c\beta \frac{d\beta }{dz},\label{eq:eqnbetaderiv}\end{equation}to transform \cref{eq:eqnmotionparkin} into a first order ODE,\begin{equation}\frac{d\beta }{dz}=\frac{A+2R}{cE_{0}}\frac{P_{4}}{D^{2}\gamma ^{3}\beta }.\label{eq:eqndirectint}\end{equation}Expressions for $P_{4}$ turn out to lead to closed-form solutions for $\beta(z)$ (and auxiliary quantities) in the lightsail temperature-limited regime and the laser power-limited regime.

\subsubsection{Lightsail Temperature-Limited Regime} \label{section:TLimitedEqns}
In the lightsail temperature-limited regime, assuming constant $A$, $P_{4}^{\prime }$ remains constant at its upper limit of $P_{4}^{\prime +}$. Also assuming constant $R$ and $E_{0}$, dimensionless range $\tilde{z}$ can be defined as
\begin{equation}\tilde{z}\equiv (A+2R) \frac{P_{4}^{\prime +}}{cE_{0}}z.\label{eq:zdimlessT}\end{equation}

\Cref{eq:einsteinpower0,eq:zdimlessT} simplify \cref{eq:eqndirectint} to a temperature-limited equation of motion,\begin{equation}\int \frac{\beta }{\left( 1-\beta ^{2}\right) ^{3/2}}d\beta =\int d\tilde{z},\label{eq:eqndirectintT}\end{equation}which directly integrates to\begin{equation}\Delta \tilde{z}=\Delta \gamma.\label{eq:eqnzT}\end{equation}This simple closed-form relation between speed and distance inverts to,\begin{equation}\beta =\sqrt{1-\frac{1}{\left( \Delta \tilde{z}+ \gamma_0\right) ^{2}}},\label{eq:eqnbetazT}\end{equation}where $\gamma_0$ is $\gamma$ at $\Delta \tilde{z}$=0.

Simple equations also relate lightsail acceleration duration $t_{s}$ to speed and distance: Defining dimensionless duration $\tilde{t}$ as\begin{equation}\tilde{t}\equiv (A+2R) \frac{P_{4}^{\prime +}}{E_{0}}t,\label{eq:tdimlessT}\end{equation}\cref{eq:eqnbeta} becomes\begin{equation}\beta =\frac{d\tilde{z}}{d\tilde{t}}.\label{eq:betanondim}\end{equation}Equating this with \cref{eq:eqnbetazT} yields\begin{equation}\int d\tilde{t}=\int \frac{d\tilde{z}}{\sqrt{1-\frac{1}{\left( \Delta \tilde{z}+\gamma_0\right) ^{2}}}},\label{eq:tintegralT}\end{equation}which integrates to\begin{equation}\tilde{t_{s}}=\Delta(\gamma \beta)\label{eq:TlimitedTime}\end{equation}given the initial condition $\Delta \tilde{z}$=$0$ at lightsail acceleration duration $\tilde{t_{s}}$=$0$. As can be illustrated on a spacetime diagram, lightsail acceleration duration relates to the radiated beam duration $t_{b}$ via\begin{equation}t_{b}=t_{s}-\frac{\Delta z}{c},\label{eq:eqntbts}\end{equation}where both durations are measured by a clock at the laser. In dimensionless form, and using \cref{eq:eqnzT}, this reduces to\begin{equation}\tilde{t}_{b}=\frac{1}{D_{0}}-\frac{1}{D}.\label{eq:eqntbtsdimless}\end{equation}

The acceleration duration $t_{s}^{\prime}$ measured by a clock on the lightsail is deduced by starting with a well-known relation for relativistic proper time $t^\prime$:\begin{equation}dt^{\prime}=\frac{dt}{\gamma(t)}.\label{eq:properTime}\end{equation}Time $t$ is substituted for $\beta$ using the equation of motion \cref{eq:eqnmotionparkin}. After integrating,\begin{equation}\tilde{t}_{s}^{\prime }=\ln \left( \frac{D}{D_{0}}\right).\end{equation}

The ratio between the time experienced by the lightsail's payload and the beam duration experienced at the laser is therefore,\begin{equation}\frac{\tilde{t}_{s}^{\prime }}{\tilde{t}_{b}}=\frac{t_{s}^{\prime }}{t_{b}}=\frac{\ln \left( \frac{D}{D_{0}}\right) }{\frac{1}{D_{0}}-\frac{1}{D}}.\end{equation}This ratio tends to $D_{0}$ in the limit $D\rightarrow D_{0}$, meaning that the payload experiences at least the duration of the accelerating beam, even if cruise velocity approaches nearly light speed.

\subsubsection{Laser Power-Limited Regime} \label{section:PLimitedEqns}
In the laser power-limited regime, $P_{3}$ remains constant at its upper limit of $P_{3}^{+}$, so that \cref{eq:P3P4} becomes\begin{equation}P_{4}=\eta _{34}(z) P_{3}^{+}.\label{eq:P1Pb}\end{equation}

The same dimensionless quantities used in the temperature-limited case can be kept by defining an efficiency-like factor,\begin{equation}\eta _{\tau}\equiv \frac{\eta _{a}P_{3}^{+}}{P_{4}^{\prime +}}\tilde{z}_{GR}.\label{eq:powerfactor}\end{equation}Assuming constant $A$, $R$, $\eta _{a}$, and $E_{0}$, and substituting \cref{eq:dickinsonefficiency,eq:dickinsontau,eq:zdimlessT,eq:P1Pb,eq:powerfactor} into \cref{eq:eqndirectint} yields,\begin{equation}\int \frac{\beta }{(1+\beta) ^{1/2}(1-\beta) ^{5/2}}d\beta =-\eta _{\tau }\int \frac{1-\tau _{d}^{2}}{\tau _{d}^{2}}d\tau _{d}.\label{eq:zintegralQ}\end{equation}The result takes the form,
\begin{equation}\eta _{\tau}I_{\tau}=I_{\beta }+K,\label{eq:zintegralQ2}\end{equation}where $K$ is a constant,\begin{equation}I_{\beta }\equiv \frac{1}{3}\left[1-\frac{(1-2\beta)( 1+\beta)^{\frac{1}{2}}}{(1-\beta)^{\frac{3}{2}}}\right],\label{eq:Ibeta}\end{equation}
and, similar to the result obtained by Rather at al. \cite{rather1976laser},\begin{equation}I_{\tau}\equiv \sqrt{\pi }\erfc(\tau_d) +\frac{1-e^{-\tau_d ^{2}}}{\tau_d}.\label{eq:Ieta}\end{equation}

Given initial lightsail position $\tilde{z}_{0}$ and speed $\beta _{0}$, $K$ is calculated using the three equations above. Given $K$ and a subsequent position $\tilde{z}_{1}$, speed $\beta _{1}$ is found by numerically solving \cref{eq:Ibeta}.

As $\tilde{z}\rightarrow \infty $, $\tau_d\rightarrow 0 $ and so $I_{\tau}\rightarrow \sqrt{\pi }$. This implies that $\beta$ tends to a finite limit $\beta _{\infty}$. Given $K$, $\beta _{\infty }$ is found by numerically solving \cref{eq:Ibeta}. This especially useful procedure shows whether or not a specified cruise velocity can be reached before attempting more involved calculations.

The acceleration duration $t_{s}^{\prime}$ measured by a clock on the lightsail is (again) deduced by starting with the well-known relation for relativistic proper time \cref{eq:properTime}. Substituting time $t$ for $\beta$ using the equation of motion \cref{eq:eqnmotionparkin} yields an integral that cannot be directly integrated as for the temperature-limited case. This integral is of somewhat similar form to the expression for $Q_{3}$, \cref{eq:Q3}, derived in the next section and can be computed using the same series solution procedure. 

Results show that a power-limited payload can experience acceleration for a shorter duration than the beam emission duration as experienced by an observer at the laser source, which is qualitatively different from the temperature-limited result that the payload experiences at least the duration of the accelerating beam. The power-limited $t_{s}$ must behave this way: As acceleration tends to zero while the beam is still on, time runs a factor of $\gamma$ slower for the lightsail payload than the laser source.

\subsubsection{Radiated Energy} \label{section:RadiatedEqns}

Energy $Q_{3}$ is radiated by the laser toward the lightsail. It determines the energy cost to accelerate each lightsail toward its destination, and it drives the laser's energy storage capacity. $Q_{3}$ is therefore essential to calculating system cost. The cost minimization procedure, given shortly, calculates $Q_{3}$ within its innermost iteration loop alongside the other trajectory equations. Therefore, we seek a faster and more accurate way to find $Q_{3}$ than numerical integration.

The essence of finding $Q_{3}$ is to integrate laser-radiated power $P_{3}(t)$ with respect to time:\begin{equation}Q_{3}=\int_{-z_{0}/c}^{t_{p}-z_{0}/c}P_{3}(t)dt=\int_{-\frac{z_{0}}{c}}^{t_{p}-\frac{z_{0}}{c}}\frac{P_{4}\left( t+\frac{z\left( t\right) }{c}\right) }{\eta _{34}\left( t+\frac{z(t)}{c}\right) }dt.\label{eq:storedenergyintegral}\end{equation}Acceleration is defined to commence when $t$=$0$, which means that the pulse begins and ends $z_{0}/c$ before it reaches the initial position of the lightsail. \Cref{sec:Nomenclature} has already described the relationship between various powers, with \cref{eq:P3P4} relating radiated power $P_{3}$ (at time $t$) to received power $P_{4}$ after a delay of $z(t)/c$. To allow the equation of motion, \cref{eq:eqnmotionparkin}, to eliminate the unknown power $P_{4}$, the origin of $t$ is first shifted by $z(t)/c$,\begin{equation}Q_{3}=\int_{0}^{t_{p}+\frac{z-z_{0}}{c}}\frac{P_{4}\left( t\right) }{\eta _{34}\left( t\right) }d\left( t-\frac{z\left( t\right) }{c}\right) =\int_{0}^{t_{p}+\frac{z-z_{0}}{c}}\frac{\left( 1-\beta \right) P_{4}\left( t\right) }{\eta _{34}\left( t\right) }dt.\end{equation}This origin shift has the effect of widening the pulse length (consistent with lightsail's acceleration duration being longer than the radiated pulse duration, as can be shown on a spacetime diagram) and reduces the power by a factor of 1-$\beta$ (as described by \cref{eq:P4pP4b}). Finally, the equation of motion \cref{eq:eqnmotionparkin} eliminates the unknown power $P_{4}$ and transforms the integral from the time domain to the velocity domain:\begin{equation}Q_{3}=\frac{E_{0}}{\eta _{a}\left( A+2R\right) }\int \frac{\gamma }{1-\beta }\frac{1}{\eta _{d}}d\beta.\label{eq:Q3}\end{equation}Defining dimensionless energy $\tilde{Q}_{3}$ as\begin{equation}\tilde{Q}_{3}\equiv Q_{3}\frac{\eta _{a}\left( A+2R\right) }{E_{0}},\label{eq:dimlessQdef}\end{equation}using \cref{eq:dickinsonefficiency}, and writing $\gamma$ in terms of $\beta$, yields a solvable energy integral:\begin{equation}\tilde{Q}_{3}=\int \frac{1}{\left( 1-\beta \right)\sqrt{1-\beta ^{2}}}\frac{1}{1-e^{-\tau _{d}^{2}\left( \beta \right) }}d\beta.\label{eq:Q3tGeneral}\end{equation}This result does not constrain any particular power to be constant and is therefore general; applicable to both temperature-limited and power-limited regimes. Choosing a regime affects the solution through $\tau _{d}(\beta)$ because $\tau _{d}$ relates to $z$ via \cref{eq:dickinsontau}, $z$ to $\tilde{z}$ via \cref{eq:zdimlessT}, and $\tilde{z}$ to $\beta$ via explicit \cref{eq:eqnzT} in the temperature-limited regime, and via implicit \cref{eq:zintegralQ2} in the power-limited regime. Unfortunately, the resulting expressions do not integrate to simple closed-form equations for either regime.

Minimum $\tilde{Q}_{3}$ occurs when $\eta _{d}$ is unity. In this special case, the indefinite integral reduces to the Doppler factor, providing a useful closed-form lower bound:\begin{equation}\tilde{Q}_{3}^{-}=\int \frac{1}{\left( 1-\beta \right) \sqrt{1-\beta ^{2}}}d\beta =\allowbreak \sqrt{\allowbreak \frac{1+\beta }{1-\beta }}=D.\end{equation}

In the general case, \cref{eq:Q3tGeneral}'s integrand is expanded into a Taylor series to find $\tilde{Q}_{3}$ without numerical integration. This series does not converge in the vicinity of non-analytic points $\beta$=1 and $\tau_{d}$=0 (corresponding to $z$=$\infty$, and to $\beta$=$\beta_{\infty}$ in the power-limited regime), but trajectories do not pass through these values anyway; they only approach them. For reasons given shortly, the series is expanded about the target speed $\beta_{1}$ and not the initial speed,\begin{equation}\Delta \tilde{Q}_{3}=\int_{\beta }^{\beta _{1}}\sum_{n=0}^{N}\frac{a_{n}(\beta _{1}) B^{n}}{n!}d\beta =\sum_{n=0}^{N}\frac{a_{n}(\beta _{1}) B^{n+1}}{\left( n+1\right) !},\label{eq:Q3tTaylor}\end{equation}where $B\equiv \beta -\beta _{1}$ is a negative quantity. The coefficient $a_{n}(\beta _{1}) $ is given by\begin{equation}a_{n}( \beta _{1}) =\lim_{\beta \rightarrow \beta _{1}}\left[ \frac{d^{n}}{d\beta ^{n}}\left( \frac{1}{1-e^{-\tau _{d}^{2}( \beta ) }}\frac{1}{( 1-\beta ) \sqrt{1-\beta ^{2}}}\right) \right].\label{eq:Q3tTayloran}\end{equation}

The system model uses algorithmic differentiation \cite{griewank2008evaluating} to compute \cref{eq:Q3tTayloran}. For this purpose, $\tau _{d}$ relates to $\tilde{z}$ via \cref{eq:dickinsontau}. Then, in the temperature-limited regime, $\tilde{z}$ relates to $\beta$ via \cref{eq:eqnzT}. Or in the power-limited regime, \cref{eq:zintegralQ2} is numerically solved to provide $\tau_{d}(\beta_{1})$, and the higher derivatives are computed from \cref{eq:zintegralQ}'s differential form combined with \cref{eq:dickinsontau}'s differential form,
\begin{equation}\frac{d\tau _{d}}{d\beta }=-\frac{1}{\eta _{\tau }}\frac{\beta }{(1-\beta) ^{2}\sqrt{1-\beta ^{2}}}\frac{\tau _{d}^{2}}{1-e^{-\tau _{d}^{2}}},\end{equation}where\begin{equation}\eta _{\tau }\equiv \frac{\eta _{a}\left( A+2R\right) P_{3}^{+}z_{GR}}{cE_{0}}.\end{equation}

The system model computes \cref{eq:Q3tTayloran} using double-precision floating-point numbers. The radius of convergence $B_{r}$ is estimated by a fairly conservative form of ratio test,\begin{equation}B_{r}=\max \left\{ \left\vert \frac{a_{0}}{a_{1}}\right\vert ,...,\left\vert \frac{a_{n-1}}{a_{n}}\right\vert \right\}.\end{equation}

Finite precision numbers clash with the exponential term $e^{-\tau _{d}^{2}( \beta )}$ to make $B_{r}$ estimates inaccurate when the Taylor series is expanded about $\beta$=0: $B_{r}$ sometimes extends beyond $\beta _{\infty }$, missing the endpoint entirely and corrupting the $\Delta \tilde{Q}_{3}$ estimate. For this reason, the series is expanded about $\beta _{1}$, the trajectory's target speed, which has a smaller $B_{r}$. 

The system model employs successive Taylor expansions (via algorithmic differentiation to 12\textsuperscript{th} order) to `hop' backwards from the target speed $\beta _{1}$ to the earlier speed $\beta$, with each hop contributing energy to $\Delta \tilde{Q}_{3}$. Hop distance $B$ is initially set to $B_{r}$. The system model considers that a hop has converged when the series' final term contributes 1 part in $10^{16}$ or less of the energy in that hop. If the hop does not converge, then $B$ is halved and the hop is retried. Retries continue until the hop converges or the model reaches a minimum hopping interval, equal to $(\beta _{1}-\beta) /10^{6}$.
 
\subsubsection{Solution Procedure}

\textit{For each} trajectory, the initial lightsail temperature (lightsail's equilibrium temperature just as acceleration commences) is calculated via a simple energy balance.

\textit{If} the initial lightsail temperature is less than the prescribed limit, then the power-limited equations of \cref{section:PLimitedEqns} are used to solve for the lightsail state at the point where the target velocity is reached.

\textit{Else if} the initial lightsail temperature is greater than or equal to the prescribed limit, then the temperature-limited equations of \cref{section:TLimitedEqns} are used instead. In this case, the model solves for the power radiated by the laser such that the lightsail equilibrium temperature equals its limit. As the lightsail accelerates away, radiated power increases to maintain the lightsail temperature at its limit despite increasing beam spillage. There comes a point where the laser reaches its maximum rated radiated power. Thereafter, the lightsail's temperature begins to fall. This transition point between the temperature-limited and power-limited equations is calculated by solving the temperature-limited equations for the distance at which the laser radiates at its maximum rated power. If the target velocity is less than the velocity at the transition point, then the temperature-limited equations are used to solve for the lightsail state as acceleration ends. Else the power-limited equations are used (with the transition point as the initial condition). In this latter case, the expended energy (needed by the cost model, for example) is then the temperature-limited energy from the start to the transition point plus the power-limited energy from the transition point to the point where the target velocity is reached.

\subsubsection{Comparison with the 2018 System Model Results}

\begin{figure*}[t]\centering\includegraphics[width=0.9\textwidth,keepaspectratio=true]{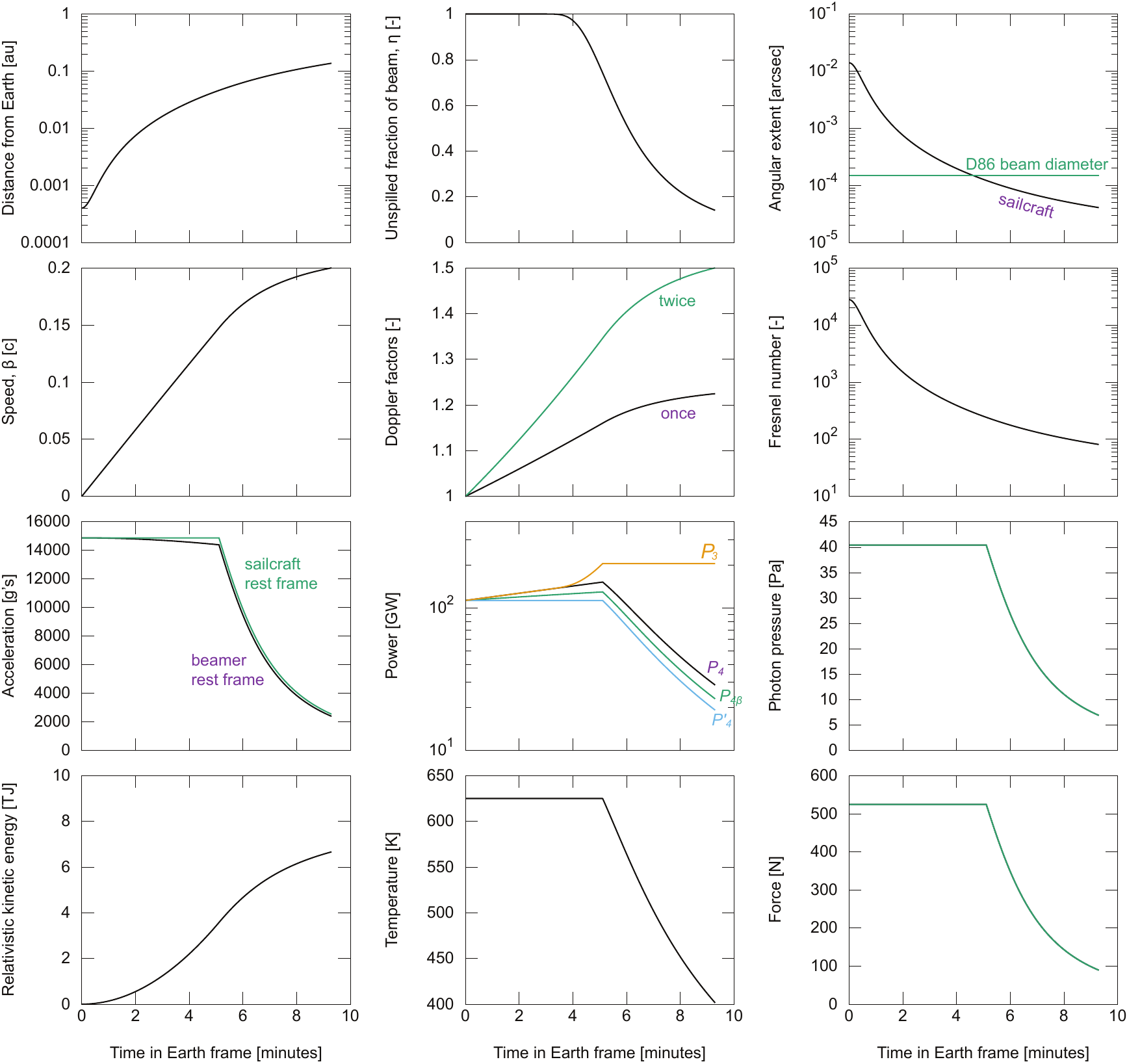}\caption{\kSpeed\ trajectory and related quantities calculated by the closed-form system model}\label{fig:02ctrajectory}\end{figure*}

The system model is now used to reproduce the \kSpeed\ trajectory that was originally produced by the 2018 system model \cite{parkin2018breakthrough}. To compare algorithms, all available input digits are given: 
\begin{itemize}\itemsep2pt
\item \qty{4.0652035386941403}{\meter} lightsail diameter
\item \qty{2.65783}{\kilo\meter} laser diameter
\item \qty{203.69892646903367}{\mega\watt} laser peak radiated power
\end{itemize}

Other constants, if needed, are those used in 2018 and repeated shortly in \cref{tab:02cinputs}, with two exceptions: First, the Stefan-Boltzmann constant has been updated in accordance with the 2019 redefinition of the SI base units and is greater than the earlier value by 1 part in $10^5$. Second, the 2018 system model code mishandles the transatmospheric propagation efficiency $\eta_{a}$, ignoring it in scaling the lightsail incident power, but including it in sizing the laser energy storage. To bypass the effect of this error, $\eta_{a}$=1 is used for computations presented within this subsection.

To produce plots with time along the x-axis, the earlier system model outputs the time and other quantities from its numerical integration steps, whereas the new system model has time as an input (not an output) and recalculates outputs that are then collected in a table. For the new model in the temperature-limited regime, time enters through \cref{eq:TlimitedTime}. In the power-limited regime, time is calculated from \cref{eq:PlimitedTime}, and the system model numerically solves for $\beta$, from which all else is calculated.

The resulting trajectory, shown in \cref{fig:02ctrajectory}, is difficult to distinguish from the corresponding plot produced by the 2018 system model \cite{parkin2018breakthrough}. The lightsail transits from temperature-limited to power-limited regimes just slightly past the \qty{5}{\minute} mark in both plots. The qualitative behavior is the same in each regime, validating the closed-form equations' correct behavior against the numerically-integrated result. 

Comparing numbers, outputs from the new code are close to the old code: \qty{0.13699}{au} vs. \qty{0.13603}{au} (0.7\% increase) in acceleration distance, \qty{80236}{\giga\joule} vs. \qty{80105}{\giga\joule} (0.2\% increase) in radiated energy, and \qty{557.94}{\second} vs. \qty{555.33}{\second} (0.5\% increase) in acceleration duration. The new expression for beam transmission efficiency differs from the old one by up to 3\% as discussed in \cref{sec:goubau}; however, it makes little difference to the result.

Acceleration distance in the new model is a function of cruise velocity and constants only, so the computed values are exact. Yet the difference between distance estimates in the two codes tends to 0.7\% as the numerical integrator's step size vanishes in the old code. The RK45 numerical integration scheme used in the old code is likely the source of the distance discrepancy and a contributing factor in the others.

\subsection{Cost Model}

The earlier cost model has been extended to include the case where power is accepted from an external grid. Pulse energy supplied in this way does not need storage, thus lowering the laser capex $C_{c}$, now given by\begin{equation}C_{c}=k_{a}\mathcal{A}_{3}+k_{l}P_{3}^{+}+k_{e}Q_{2}.\label{eq:eqncost}\end{equation}

Laser area $\mathcal{A}_{3}$ and peak radiated power $P_{3}^{+}$ are dependent variables of the system model. Factors $k_{a}$, $k_{l}$, and $k_{e}$ are independent user-supplied values for cost per unit area, cost per unit power, and cost per unit energy stored; they are technology figures of merit. Stored pulse energy $Q_{2}$ is modified to account for maximum available electric transmission line power $P_{1}^{+}$,\begin{equation}
Q_{2} =
  \begin{cases}
    0 & \text{if $\frac{P_{3}^{+}}{P_{1}^{+}}<\eta _{13}$}\\
    \frac{Q_{3}}{\eta _{23}}-\eta _{12}P_{1}^{+}t_{p} & \text{if $\frac{P_{3}^{+}}{P_{1}^{+}}\geq \eta _{13},\frac{P_{4}^{\prime +}}{P_{1}^{+}}>\eta _{14,t=0}$}\\
    \frac{\Delta Q_{g}}{\eta _{23}}-\eta _{12}P_{1}^{+}\Delta t_{g} & \text{otherwise,}
  \end{cases}
\label{eq:eqnQ1}\end{equation}where $\eta _{13}\equiv\eta _{12}\eta _{23}$ and $\eta _{14}\equiv\eta _{13}\eta _{35}$.

If there is sufficient power from the grid ($P_{3}^{+}$\textless$\eta _{13}P_{1}^{+}$), then no energy storage is needed and the lightsail is accelerated by grid power alone. Else if grid power initially heats the sail to its maximum temperature ($P_{4}^{\prime +}$\textgreater$\eta_{14,t=0}P_{1}^{+}$), then the power supplied by the grid is constant at its maximum $P_{1}^{+}$ and can be multiplied by the pulse duration $t_{p}$ to obtain the pulse energy supplied by the grid, $\eta _{12}P_{1}^{+}t_{p}$. Storage is therefore required for the radiated energy $Q_{3}/\eta _{23}$ minus that supplied by the grid. Otherwise, the lightsail is initially power-limited. In this case, no stored energy is needed until time $t_{g}$, after which any energy in excess of that supplied by the grid is instead supplied from storage. This excess energy is defined as $\Delta Q_{g}\equiv Q_{3}-Q_{3g}$, the energy radiated up to time $t_{p}$ minus the energy radiated up to $t_{g}$. From time $t_{g}$ to the pulse's cutoff at time $t_{p}$, the grid supplies its maximum $P_{1}^{+}$, so $P_{1}^{+}(t_{p}-t_{g})$ is subtracted from the required energy storage. $Q_{3g}$ and $t_{g}$ are obtained by solving the temperature-limited equations of \cref{section:TLimitedEqns} such that $P_{1}$=$P_{1}^{+}$.

Opex $C_{o}$ is approximated to pulse energy cost, and pulse energy originates from the grid whether or not it is stored in an intermediate step:\begin{equation}C_{o}=k_{g}Q_{1}.\label{eq:eqnCo}\end{equation}

These equations imply that efficiencies $\eta _{12}$ and $\eta _{23}$ do not depend on whether radiated energy $Q_{3}$ is supplied directly from the grid or via intermediate storage. This approximation can be revisited in future.

\subsection{Cost Minimization Procedure}

\begin{figure}[hbtp]\centering\includegraphics[width=\columnwidth]{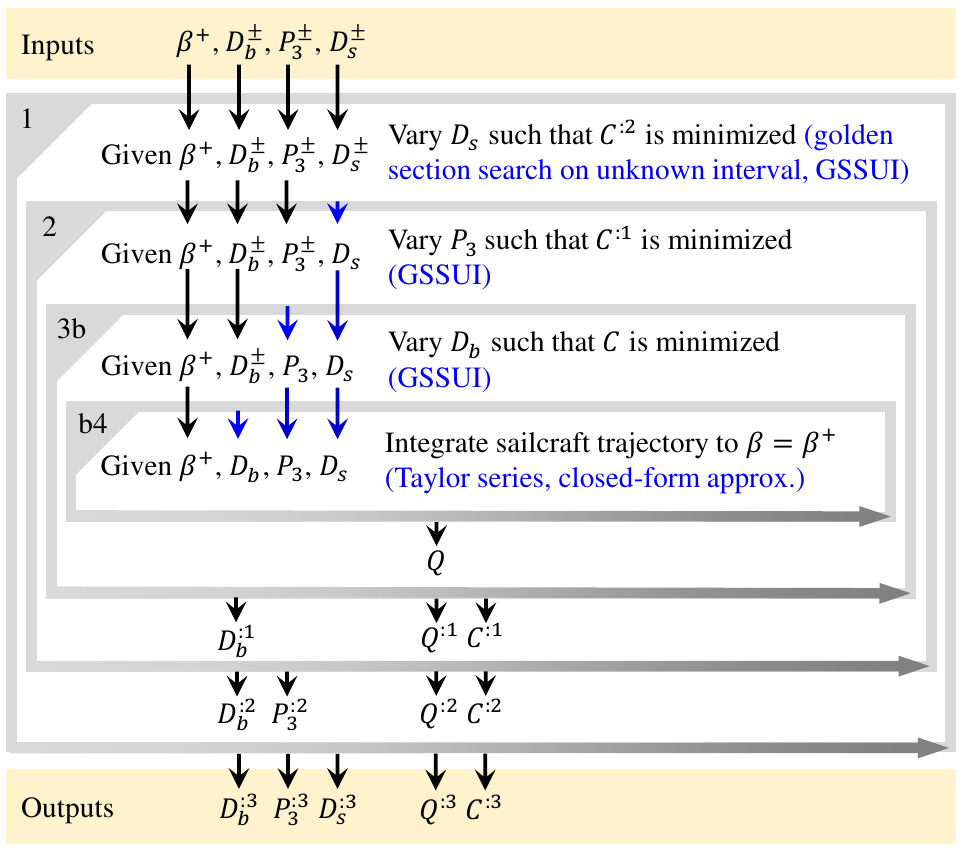}\caption{Solution procedure}\label{fig:solutionproc}\end{figure}

As before, the system model uses nested optimizations to ensure that the lightsail reaches its specified cruise velocity and that all system elements are sized to minimize cost. This new solution procedure, shown in \cref{fig:solutionproc}, no longer explicitly calculates the minimum laser diameter for which cruise velocity is reachable (thus deleting former iterations 3a and a4). Instead, the golden-section search logic has been modified to accept an additional boolean value (cruise velocity reached) that signals whether the function being minimized (cost) is valid for this input value. This means that if cost is evaluated for a given point design, but that point design does not turn out to reach the specified cruise velocity, then the search region shrinks accordingly, and optimization continues thereafter.

Thus far, the deliberately vague word `cost' has been used because the minimized quantity has changed from capex to,\begin{equation}C\equiv C_{c}+n_{o}C_{o}\label{eq:C}\end{equation}instead, where $n_{o}$ is the number of lightsail accelerations to minimize the cost for, $C_{o}$ is the single-acceleration opex, and $C_{c}$ is the capex. This change is made because minimizing capex alone gives too little regard to opex, which can be very high for smaller grid-dominant missions (see \cref{section:GridAugmented}). Specifically, the optimizer shrinks the laser, reducing capex but potentially increasing opex by decreasing energy transfer efficiency. Minimizing $C$ instead of $C_{c}$ ensures that the optimizer tries to balance capex and opex.

\section{System Performance Maps}

System performance maps show how system-level figures of merit (and other engineering quantities) vary with cruise velocity and payload mass. Each plotted point corresponds to a point design computed by the system model. It is the intent of these maps to provide a sense of the cost, power, and physical scales involved at the earliest stage of conceptualizing new missions. With this in mind, the lightest payload in the map is \kpMapsPayloadMin, enough to carry $10^{5}$ human cells. The heaviest is \kpMapsPayloadMax, enough to carry 225 International Space Stations. The slowest cruise velocity in the map is \qty{.0001}{c}, which is less than the record \qty{.0005}{c} reached by the Parker Solar Probe (via conventional propulsion with gravity assists). The fastest is \qty{.99}{c}, at which speed the \qty{4.4}{ly} journey to Alpha Centauri, for example, is experienced as \qty{225}{\day} in the lightsail's rest frame.

Other constants used in computing the maps are summarized in \cref{tab:02cinputs}. These values are largely unchanged from before \cite{parkin2018breakthrough}, so the descriptions and justifications are not repeated here. Only the former ``wallplug to laser efficiency'' is changed; it is decomposed into $\eta _{12}$ and $\eta _{23}$ because the cost model now requires this distinction. $\eta _{12}$ is chosen such that the ``wallplug to laser efficiency'' is unchanged, allowing comparisons with earlier results.
  
\begin{table}[h]\caption{System model constants}\label{tab:02cinputs}\centering\begin{tabular}{l}\hline			
\kpWavelength\ wavelength \\
\qty{60000}{\kilo\meter} initial sail displacement from laser\\
\\
\qty{0.2}{\gram\per\meter\squared} areal density \\
\kpAbsorptance\ spectral normal absorptance at \kpWavelength\\
70\% spectral normal reflectance at \kpWavelength\\
\kpMaxTemp\ maximum temperature\\
\kpEmittance\ total hemispherical emittance (2-sided, \kpMaxTemp)\\
\\
\$\qty{0.01}{\per\watt} laser source cost ($k_{l}$)\\
\$\qty{500}{\per\meter\squared} optics cost ($k_{a}$)\\
\$\qty{50}{\per\KWH} storage cost ($k_{s}$)\\
\$\qty{0.1}{\per\KWH} grid energy cost ($k_{g}$)\\
100\% grid to storage efficiency ($\eta _{12}$)\\
50\% storage to laser efficiency ($\eta _{23}$)\\
70\% transatmospheric propagation efficiency ($\eta _{a}$)\\
\\
100 operations included in cost minimization ($n_{o}$)\\
\hline\end{tabular}\end{table}

\subsection{Stored Energy Only}

Asserting zero grid input (in addition to the inputs given in \cref{tab:02cinputs}) yields the map shown in \cref{fig:MapNoGrid}. As would be expected, the laser's capex (plot [1,3] where [r,c] denotes the row and column number, with [1,1] corresponding to the top left) grows as the payload gets heavier, cruise speeds up, or any combination thereof. Maximum capex is set at \$10T because Flyvbjerg \cite{flyvbjerg2014you} estimates worldwide annual megaproject spending to be \$6T to \$9T circa 2014. Greater capex than that is colored in magenta, whereas less than \$1M minimum is colored in cyan. 

In general, magenta signals greater than maximum and cyan signals less than minimum throughout the maps. Undesirable limits tend toward the color red in the plots, and desirable limits tend toward blue. Intermediate values are colored in gray and white because the two neutral tones show more detail.

Plots for which the limits are neither good nor bad use black as one bound and white as the other. In this case, intermediate values are plotted in blue because this highlights details that would otherwise be lost in gray.

\begin{figure*}[p]\centering\includegraphics[width=1.0\textwidth,keepaspectratio=true]{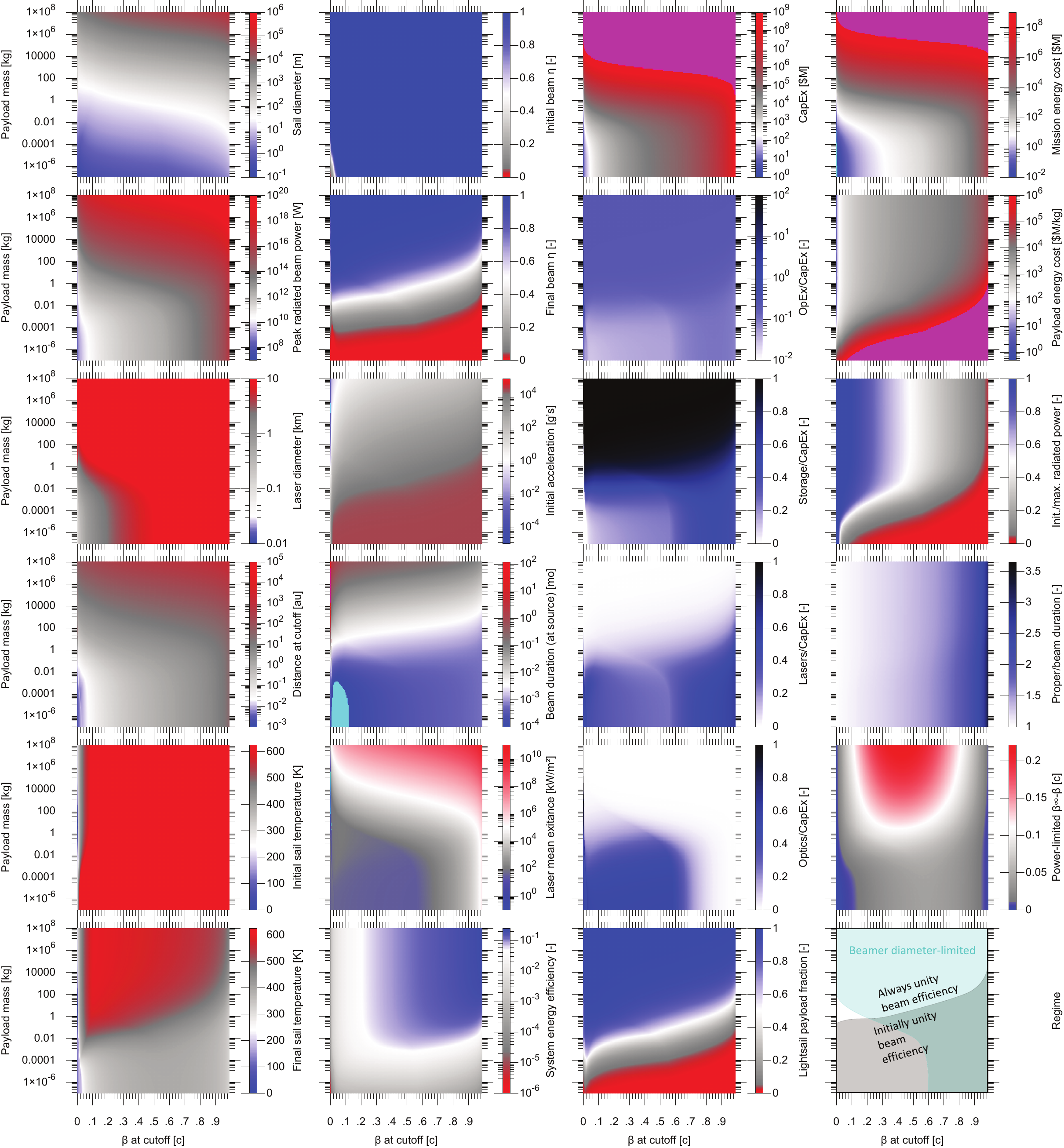}
\caption{System performance map for a lightsail that is accelerated by a laser whose pulse energy is entirely stored locally to the laser. All x-axes are $\beta $ at cutoff (equal to cruise velocity) and are plotted on a linear scale. All y-axes are payload mass and are plotted on a logarithmic scale.}\label{fig:MapNoGridNonLog}\end{figure*}

\begin{figure*}[p]\centering\includegraphics[width=1.0\textwidth,keepaspectratio=true]{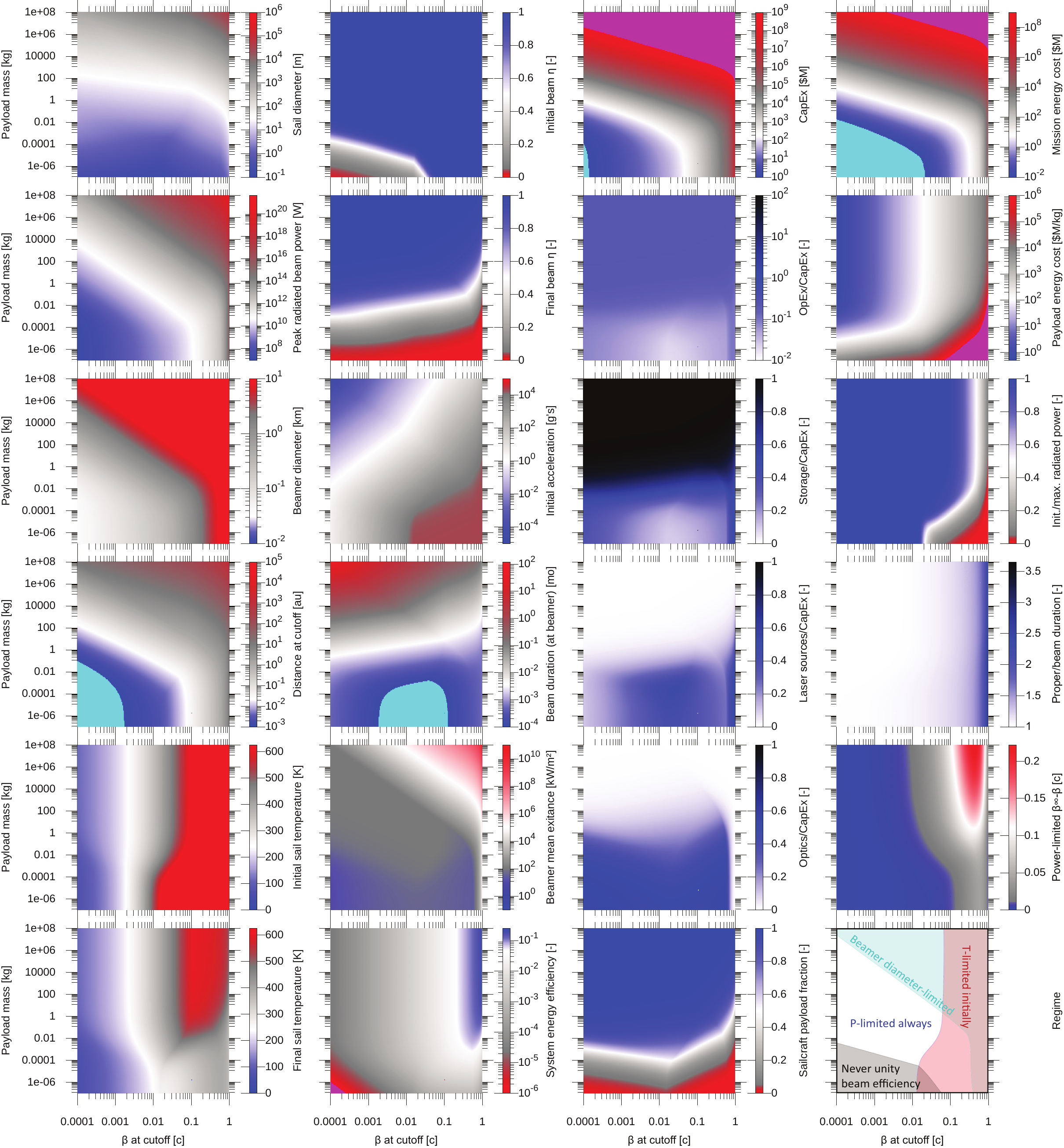}
\caption{System performance map for a lightsail that is accelerated by a laser whose pulse energy is all stored locally to the laser. All x-axes are $\beta $ at cutoff (equal to cruise velocity) and are plotted on a logarithmic scale. All y-axes are payload mass and are plotted on a logarithmic scale.}\label{fig:MapNoGrid}\end{figure*}

\begin{figure*}[p]\centering\includegraphics[width=1.0\textwidth,keepaspectratio=true]{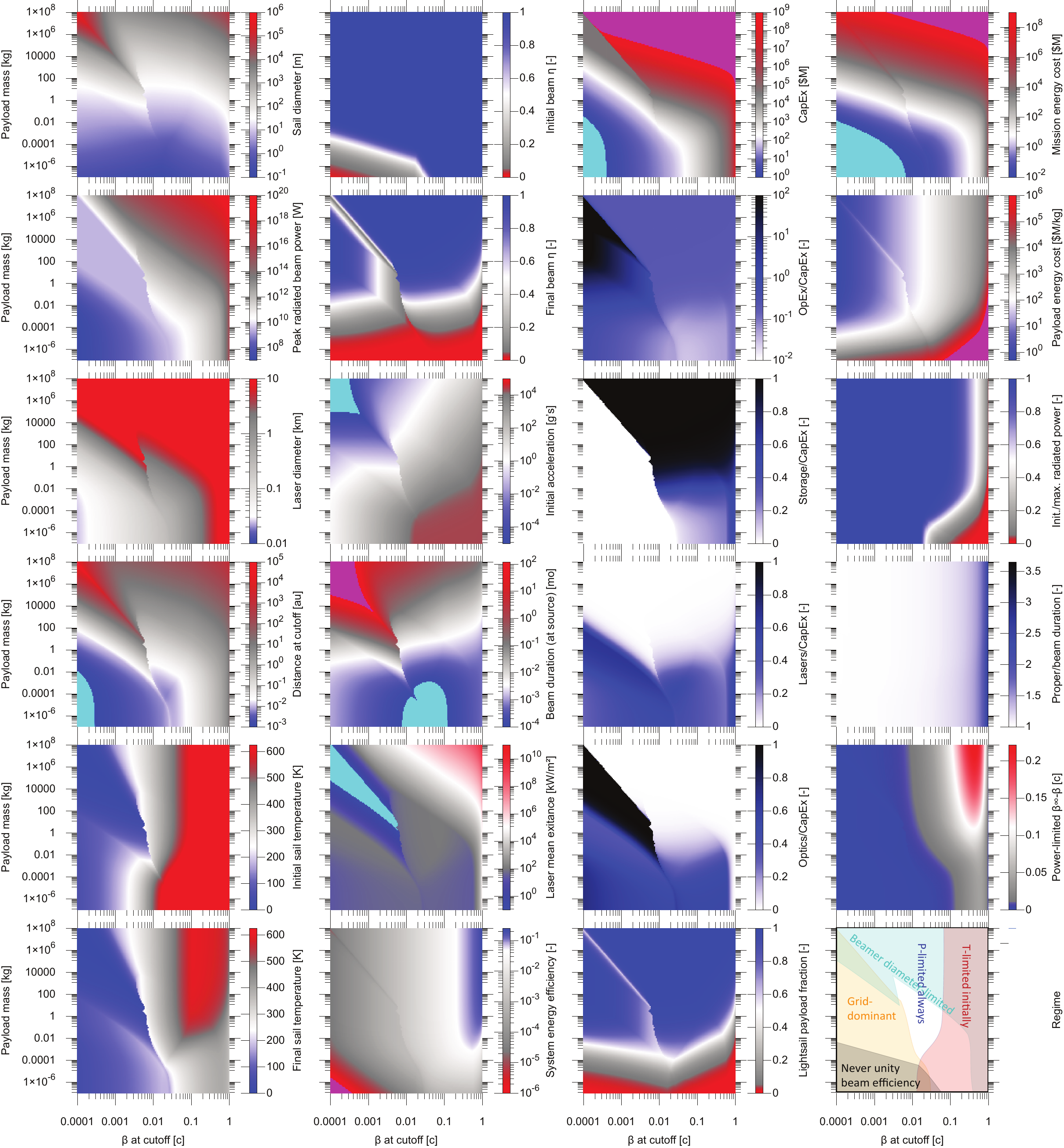}
\caption{System performance map for a lightsail that is accelerated by a laser whose pulse energy is augmented by transmission line to a regional or national grid. All x-axes are $\beta $ at cutoff (equal to cruise velocity) and are plotted on a logarithmic scale. All y-axes are payload mass and are plotted on a logarithmic scale.}\label{fig:MapGrid}\end{figure*}

The map's dynamics can be reasoned by starting with the primary quantities obtained from iterations 1, 2, and 3b in \cref{fig:solutionproc}: Sail diameter, laser power, and laser diameter respectively. The corresponding maps are plots [1,1], [2,1], and [3,1] in \cref{fig:MapNoGrid}. Sail diameter and peak radiated beam power are unconstrained, whereas laser diameter is maximally \kpMapsLaserDiaMax. When the optimizer (iterations 1, 2, and 3b) reaches this limit, the limit becomes an immovable obstacle to lowering cost, and therefore a driving constraint. 

When laser diameter is maximal, the optimizer varies sail diameter and peak radiated beam power instead, and in this way the solution's dynamics qualitatively change within the uniform red region of plot [3,1]. For example, the laser's mean radiant exitance (plot [5,2]) tends to be between \qtyrange{1}{1000}{\kilo\watt\per\meter\squared} until maximum laser diameter is reached, then rapidly intensifies a thousand-fold (because area is constrained). This more intense region reddens in the radiant exitance plot, signaling undesirability. Ultimately, physics limits radiant exitance to \qty{e6}{\kilo\watt\per\meter\squared} for ground-based lasers because the surrounding air breaks down into a plasma \cite{raizer1980optical}, and to \qty{e22}{\kilo\watt\per\meter\squared} for space-based lasers because free space transparency is limited by electron-positron pair production \cite{bell2008possibility}.

Another example of how constraints affect map dynamics is the sail temperature, which is limited to \kpMaxTemp. \Cref{fig:MapNoGrid}'s bottom-left plot and the plot above it show the sail temperature at the start and finish of its acceleration. For missions spanning a wide range of cruise velocities and payload masses, sail temperature is never a constraint. Whereas in other regions, temperature is a constraint at acceleration's start (plot [5.1]). When sail temperature becomes a constraint, it is not directly optimized as laser diameter is, but instead determines whether the lightsail's trajectory is modeled by the temperature-limited equations in \cref{section:TLimitedEqns} or the power-limited equations in \cref{section:PLimitedEqns}.

The impossibility of better than perfect energy transfer acts as a boundary condition imposed by physics itself. \Cref{fig:MapNoGrid} shows beam transmission efficiency as acceleration begins (plot [1,2]) and ends (plot [2,2]). For heavy enough payloads, efficiency is unity throughout. In regions where beam transmission efficiency has reached unity, one might wonder why the optimizer still chooses the maximum laser diameter in some cases? A smaller laser would reduce capex without affecting the efficiency. However, capex (essentially all energy storage cost in this regime) is not the optimized quantity. The optimized quantity is given by \cref{eq:C}, so pulse energy cost rules in this regime. Doubling the laser diameter halves the sail diameter, in turn reducing sail mass, in turn reducing lightsail energy. Therefore, the optimizer has a different reason to maximize laser diameter, even in cases where the payload fraction (plot [6,3]) approaches unity.

A summary of regimes (plot [6,4]) combines the key constraints of maximum laser diameter, maximum sail temperature, and perfect beam transmission efficiency. Each regime is driven slightly differently, so the solution behaves slightly differently within it. Piecing together these various behaviors is a way to understand and verify the maps.

Beam pulse duration (measured at the laser in plot [4,2]) varies over \qtyrange{1}{e7}{\minute} (\qtyrange{2e-6}{20}{\year}). Any trajectory lasting more than a few minutes will have to contend with orbital obstacles and the Earth's rotation relative to a lightsail's intended destination. Beam interruption or handoff to other lasers will require the beam to re-entrain the lightsail.

Surprisingly, system energy efficiency (plot [6,2]) climbs with lightsail cruise velocity, tending toward 22\% as missions aim for relativistic speeds. This high-efficiency region somewhat corresponds to the region in which the sail is most temperature-limited (the red region in plot [6,1]).

Opex/capex ratio (plot [2,3]) is $n_oC_o/C_c$ (not $C_{o}/C_{c}$) and might therefore be expected to be $\mathcal{O}(1)$ after cost optimization. However, the ratio is ${\sim}1/5$ for missions carrying more than \qty{10}{\gram} of payload. For lighter missions, the ratio decreases further, approaching ${\sim}1/100$ minimum at around \qty{0.02}{c} cruise velocity. Therefore, it is incorrect to assume that capex and opex (total for 100 missions) are roughly equal for cost-optimal point designs, as has sometimes been used as a rule of thumb.

Energy storage (plot [3,3]) is the dominant capex for missions with heavier payloads than \qty{10}{\gram}; laser source (plot [4,3]) and optics (plot [5,3]) expenses prove negligible. For missions with lighter payloads than \qty{10}{\gram}, storage, laser source, and optics expenses all matter, with optics being roughly half the capex except for relativistic cruise velocities. Laser sources trade off against storage for the other half of the capex. Laser source cost is least important for missions having around \qty{0.5}{c} cruise velocity, and also for slower missions having \qty{0.001}{c} cruise velocity or less.

Lightsails carrying heavier payloads than \qty{10}{\gram} are mostly payload mass (plot [6,3] shows lightsail payload fraction) until their cruise velocity approaches light speed. When payload mass dominates, the sail may yet have greater extent than the payload, or it may be regarded as an optical layer that coats the payload. In the latter case, the payload would have at least the diameter required of the sail. For example, if the lightsail were a gas-filled balloon containing a garden, then the sail might be regarded as a special outer coating on the balloon, and all inside would be regarded as the payload. This example neglects the problematic dynamics of balloons \cite{atwater2018materials}, and it neglects provisions that deflect the transmitted fraction of the beam away from the absorbing parts of the payload during acceleration.

System-level performance in the relativistic limit is more easily discerned by plotting the cruise velocity on a linear scale, as done in \cref{fig:MapNoGridNonLog}. The lightsail payload fraction (plot [6,3] as before) is now seen to decline as cruise velocity increases. Payload energy cost (plot [2,4]) starts to climb into the \$B/kg range despite (or thanks to) the region of high system energy efficiency, which on a linear scale is broad and occupies a great deal of the mission design space. Though it is hard to tell from plot [4,4], the ratio of beam duration measured on board the lightsail relative to that measured at the laser is a weak function of payload mass.

Previously, it had been found that it is cost optimal to oversize the laser's maximum radiated power, ramping up from low initial power to compensate for increasing losses as lightsails accelerate away (eventually saturating at a maximum value). The initial to maximum radiated power ratio (plot [3,4]) shows that this is not universally the way to save money: For slower cruise velocities than \qty{0.02}{c}, it is always cheapest for the laser to radiate maximum power throughout acceleration. For cruise velocities faster than \qty{.2}{c}, it is always cheapest to oversize laser power. And for cruise velocities between, it depends on the payload mass. The regimes summarized in plot [6,4] help to see that cost favors overpowered lasers when the lightsail is initially temperature-limited.

\subsection{Stored Energy Augmented by Grid} \label{section:GridAugmented}

The prior section has predicted that energy storage capex dominates laser capex for missions carrying heavier payloads than \qty{10}{\gram} (as shown in \cref{fig:MapNoGridNonLog} plot [4,3]). To lower the energy storage cost, power drawn from laser storage can be augmented by a regional or national grid\footnote{This obvious improvement was not originally considered because it would have made negligible difference to the Centauri system mission cost.}, thus reducing the energy storage needed at the laser site.

Grid power is limited by the transmission line's capacity and the generating capacity available via that grid. At this time, grids in many countries routinely transport multi-gigawatt power levels, so this performance map assumes that up to \kpMapsGridP\ is available to the laser for an indefinite period. Transmission line capex is not included in the system cost, nor is the electrical substation's cost; these costs are regarded as being amortized into the \$\qty{0.1}{\per\KWH} grid energy rate.

Grid-connected maps are shown in \cref{fig:MapGrid} (the page after non-grid maps in \cref{fig:MapNoGrid}, to make comparisons easier). Grid augmentation creates a new solution branch that capitalizes on the grid connection, changing other system specifications to maximize this advantage. When the payload and speed are low enough, this grid-dominant branch supplies the lowest capex, and when they are high enough, the grid connection fades into insignificance, leaving the solution almost as it was without grid input. The maps show a discrete dividing line, on the left side of which the optimizer chooses the grid-dominant solution branch, and on the right side of which the optimizer chooses solutions that are essentially unchanged from the previous map.

The full dividing line between solution branches\footnote{Because the right side of the map is virtually unchanged, the linear map of \cref{fig:MapNoGridNonLog} can be used in this case as well, with all visible changes being in the leftmost few columns of pixels.} is most easily discernible in the opex to capex ratio (plot [2,3]). The left side of this plot shows opex exceeding capex, meaning the laser becomes cheaper to build than to operate 100 times.

As the dividing line is approached from the grid-dominant (lower left) side, the optimizer wrings all it can from the power-limited transmission line, rapidly increasing the laser diameter (plot [1,1]) leading to the thin gray outline in plot [6,3] as the lightsail's payload fraction drops. There is a wobbliness to the dividing line in some sections. Nevertheless, all plotted points are valid point designs. Also, the capex (plot [1,5]) and opex (plot [2,5]) of the grid-dominant region compared to those in \cref{fig:MapGrid} and \cref{fig:MapNoGrid} show that the optimizer is not picking noticeably more expensive solutions than the storage-only map. If the map chooses the wrong minimum, it chooses a valid mission point design that is not as cheap as it could be.

Grid input changes which laser subsystems drive capex. Referring to \cref{fig:MapNoGrid} once again, storage cost (plot [3,3]) drops to zero within the grid-dominant region, leaving optics and laser sources as the two major expenses. Only within the grid-dominant region's upper part does optics dominate the capex (plot [5,3]). As the peak radiated beam power (plot [2,1]) falls below \qty{2.5}{\giga\watt}, laser source capex (plot [4,3]) becomes the largest share. Continuing down in payload mass, optics cost once again becomes more important as beam transmission efficiency begins to drop (plots [1,2] and [2,2]).

\section{Illustrative Point Designs}

Grid input reduces mission capex by 1-5 orders of magnitude, with 1-3 orders of magnitude being more typical. The biggest savings are for slow and heavy missions. For example, a \pAMass\ payload \pASpeedc\ mission capex collapses from \pAGCapex\ to \pAOCapex, as shown in \cref{tab:10kgpointdesigns}. This 40-fold drop occurs while the pulse energy cost and system energy efficiency remain comparable in both cases. The optimizer chooses \qty{2.5}{\giga\watt} peak radiated power from the laser (as it does over a sizable region of plot [2,1] of \cref{fig:MapNoGrid}) because that corresponds to the \kpMapsGridP\ power limit of the grid transmission line. Laser source cost is therefore fixed at \$25M for this solution family. The lightsail's mass and speed fix the kinetic energy, which in turn drive the pulse duration for an effectively-fixed beam power. Because the pulse takes 15 times longer to accelerate the lightsail, it gets farther away before cruise velocity is attained. To avoid greater beam spillage, which is expensive, the optimizer quadruples the sail diameter and doubles the laser diameter. 

\begin{table}[h]\caption{\pAMass\ payload \pASpeedc\ (\pASpeedau) mission point design}\label{tab:10kgpointdesigns}\centering\begin{tabular}{lcc}\hline
&No grid input&Grid input\\\hline 
Capex&\pAGCapex&\pAOCapex\\
 - of which laser sources&\$320M&\$25M\\
 - of which optics&\$160M&\$590M\\
 - of which storage&\$26B&\$0\\
Opex per mission&\$51M&\pAOOpex\\
Sys. energy efficiency&0.025\%&0.024\%\\
Sail diameter&\qty{18}{\meter}&\pAOSIIDia\\
Laser diameter&\qty{0.64}{\kilo\meter}&\pAOSIDia\\
Peak radiated power&\qty{32}{\giga\watt}&\pAOPower\\
Pulse duration&\qty{8.1}{\hour}&\qty{120}{\hour}\\
Initial acceleration&\qty{1.0}{g's}&\pAOAccnInit\\
\hline\end{tabular}\end{table}

\begin{table}[h]\caption{\pBMass\ payload \pBSpeedc\ (\pBSpeedau) mission point design}\label{tab:10kgpointdesigns10xslow}\centering\begin{tabular}{lcc}\hline
&No grid input&Grid input\\\hline 
Capex&\pBGCapex&\pBOCapex\\
 - of which laser sources&\$18M&\$11M\\
 - of which optics&\$9.7M&\pBOOpex\\
 - of which storage&\$2.6B&\$0\\
Opex per mission&\$5.1M&\$5.2M\\
Sys. energy efficiency&0.0025\%&0.0024\%\\
Sail diameter&\qty{14}{\meter}&\qty{26}{\meter}\\
Laser diameter&\qty{160}{\meter}&\qty{120}{\meter}\\
Peak radiated power&\qty{1.8}{\giga\watt}&\qty{1.1}{\giga\watt}\\
Pulse duration&\qty{14}{\hour}&\qty{24}{\hour}\\
Initial acceleration&\qty{0.06}{g's}&\qty{0.04}{g's}\\
\hline\end{tabular}\end{table}

\begin{table}[h]\caption{\pCMass\ payload \pCSpeedc\ mission point designs}\label{tab:01mgpointdesigns}\centering\begin{tabular}{lcc}\hline
&No grid input&Grid input\\\hline 
Capex&\pCGCapex&\pCOCapex\\
 - of which laser sources&\$9.8M&\$9.9M\\
 - of which optics&\$13M&\$11M\\
 - of which storage&\$2.8M&\$0\\
Opex per mission&\$5.5k&\pCOOpex\\
Capex/payload&\$250t/kg&\$210t/kg\\
Opex/payload&\$55B/kg&\$120B/kg\\
Sys. energy efficiency&0.0089\%&0.0046\%\\
Sail diameter&\qty{16}{\centi\meter}&\pCOSIIDia\\
Laser diameter&\qty{180}{\meter}&\qty{170}{\meter}\\
Peak radiated power&\qty{980}{\mega\watt}&\qty{990}{\mega\watt}\\
Pulse duration&\qty{1.7}{\min}&\qty{3.7}{\min}\\
Initial acceleration&\qty{9500}{g's}&\qty{8300}{g's}\\
\hline\end{tabular}\end{table}

\begin{table}[h]\caption{\pDMass\ payload \pDSpeedc\ (\pDSpeedau) mission point designs}\label{tab:100ktpointdesigns}\centering\begin{tabular}{lcc}\hline
&No grid input&Grid input\\\hline 
Capex/payload&\pDGCapex&\pDOCapex\\
Opex/payload&\pDGOpex&\pDOOpex\\
Sys. energy efficiency&1.7\%&1.7\%\\
Sail diameter&\qty{7.4}{\kilo\meter}&\pDOSIIDia\\
Laser diameter&\qty{10}{\kilo\meter}&\qty{10}{\kilo\meter}\\
Peak radiated power&\qty{380}{\peta\watt}&\pDOPower\\
Pulse duration&\qty{20}{\day}&\pDODurPul\\
Acceleration duration&\qty{21}{\day}&\pDODurAccn\\
Initial acceleration&\qty{1.3}{g's}&\qty{1.3}{g's}\\
Final acceleration&\qty{1.1}{g's}&\qty{1.1}{g's}\\
\hline\end{tabular}\end{table}

For the same \qty{10}{\kilo\gram} payload, slowing cruise velocity from \qty{.001}{c} to \qty{.0001}{c} cuts the mission capex 36-fold to \pBOCapex\ and the opex 10-fold to \pBOOpex, as shown in \cref{tab:10kgpointdesigns10xslow}. At this scale, the R\&D needed to attain the assumed technology figures of merit is likely more expensive than the laser capex. But, once the figures of merit are attained, subsequent lasers will cost \pBOCapex, and the capex should continue to decline with each further facility as described by an experience curve. The \qty{24}{\hour} pulse duration is fine for missions that head near the direction of the north or south celestial poles, but missions heading into the ecliptic plane will see their accelerating laser set over the horizon, require more than one accelerating laser with handover from one to the next.

It is worth mentioning that roughly the same laser can also be used to accelerate a much lighter lightsail (\qty{0.1}{\milli\gram}) to a much higher cruise velocity (\qty{.01}{c}), as shown in \cref{tab:01mgpointdesigns}. This payload is the lightest considered in this work and is about the mass of 100,000 human cells, or a microbiome. The \qty{4}{\min} acceleration time is fast enough that a single laser can accelerate many such lightsails per day to destinations in the ecliptic plane or outside it.

The Centauri system mission accelerates at \qty{14900}{g's}, but the acceleration map (plot [3,2] of \cref{fig:MapNoGrid}) shows that other mission accelerations are rarely higher (\qty{20600}{g's} at most), and often far lower: \Cref{tab:10kgpointdesigns} shows the \qty{10}{\kilo\gram} payload being accelerated very gently at \pAOAccnInit. Interestingly, the white line extending from the left of plot [3,2] diagonally up to the top represents missions with \qty{1}{g} initial acceleration. The heaviest such mission has \pDMass\ payload. As described in \cref{tab:100ktpointdesigns}, this accelerates to \pDSpeedc\ over the course of \pDODurPul\ (plot [4,2]) using a \pDOSIIDia\ diameter sail (plot [1,1]), arriving at the Centauri system \qty{60}{\year} later; within a human lifetime. At today's energy prices, the payload energy cost (plot [2,4]) is \pDGOpex. Faster travel only ever increases this. Also, raising payload mass above \qty{1}{\kilo\gram} barely increases \$/kg. Long before this, materials R\&D would improve these figures of merit by increasing sail reflectance and decreasing sail absorptance relative to the values assumed here.

\section{Conclusions}

Closed-form trajectory equations have been derived for laser-driven lightsails and used in place of numerical integration in the Starshot system model. Consequently, point designs are computed 1-2 orders of magnitude faster. The rate-determining step is now to compute the total energy radiated by the laser over the duration needed to accelerate the lightsail: The energy integral has been replaced with a series approximation by transforming the integrand into a function of lightsail speed, then Taylor expanding the resulting expression. Twelve terms of the series are computed via algorithmic differentiation. The series is recomputed many times so as not to exceed the convergence radius, so in this sense it does step like a numerical integrator, though with fewer steps. 

The speed and accuracy of the new scheme surpass the RK45 numerical integrator used in 2018: When comparing the Centauri system mission trajectory, the new acceleration distance is 0.7\% longer, the radiated energy is 0.2\% greater, and the acceleration takes 0.5\% longer. Acceleration distance is now a function of cruise velocity and constants only, so the computed values are exact. Therefore, the old code's RK45 numerical integrator is responsible for most of the remaining difference, together with other factors like the simplified efficiency formula and the 2019 redefinition of SI base units, which noticeably alters the Stefan-Boltzmann constant. Also, the new code has a finite but smaller error in its energy and time estimates resulting from the finite number of Taylor series terms and the double-precision floating-point representation.

Crucially, the maximum and minimum values of numerically-solved variables no longer need to be manually adjusted, allowing point designs to converge without user intervention over an expanded design space that includes missions having \kpMapsPayloadMin\ to \kpMapsPayloadMax\ payload and \kpMapsCruiseRangeC\ (\kpMapsCruiseRangeAuYr) cruise velocity. Speed and autonomy have combined to enable work to progress from cost-optimized point designs to whole performance maps. These maps reveal differing solution regimes characterized by their driving constraints.

Performance maps have been generated for the case in which the laser draws all power from its on-site energy storage system, and for the case in which the laser draws up to \kpMapsGridP\ directly from the grid (to augment the power drawn from on-site energy storage). Direct grid power introduces a new solution branch that matches laser power to transmission line capacity by increasing sail diameter and accelerating slower. The performance maps show a discrete dividing line (and not a continuous transition) to the left side of which the optimizer chooses the grid-dominant branch, and to the right side of which the optimizer ignores it. The wobbliness of the dividing line for lighter payloads than \qty{100}{\kilo\gram} is likely a numerical artifact caused by the optimizer, a nested golden-section search. This type of optimizer assumes a cost function with one minimum, yet may be seeing two (one corresponding to each solution branch). Despite this, all plotted points are valid point designs that reach the target cruise velocity, and the optimizer demonstrably picks close-to-optimal solutions near the dividing line. In future, the optimizer can be upgraded to choose the lowest minimum in the vicinity of the dividing line, and this could lower costs over a wider range of missions.

Solar system through to interstellar precursor missions are relatively slow and heavy, and the laser needed to accelerate them was disappointingly expensive due to the cost of on-site energy storage. Introducing direct grid power has made these lasers affordable; they are 5 orders of magnitude cheaper in some cases, with 1-3 orders of magnitude being more typical. For example, direct grid power collapses capex from \pAGCapex\ to \pAOCapex\ for a \pAMass\ payload to \qty{.001}{c} (\qty{63}{au\per\year}). Such a mission could be a Solar system cubesat that reaches Neptune (\qty{30}{au}) in 6 months, and/or the Solar Gravitational Lens threshold (\qty{550}{au}) in less than a decade. Accelerating the cubesat would take \pAOOpex\ worth of pulse energy per mission (assuming \$\qty{0.1}{\per\KWH} grid energy cost). If the mission were to produce a gravitational map by using 100 cubesats, then the \$6B energy cost would be in the ballpark of traditional flagship mission costs (and of course, further R\&D could reduce energy requirements). A gravitational mapping effort on the 100+ cubesat scale would ideally include sensors to build a map of the heliopause and the dust environment beyond it. These cubesats would not require imaging sensors or precision trajectories, nor return a large volume of data.

For small missions, the lowest capital costs will be attained by lasers that are adjacent to existing transmission lines with adequate capacity and connected to low-cost power markets. Transmission line capex has not yet been explicitly included in the system cost, nor has electrical substation capex; these have been regarded as being amortized into the \$\qty{0.1}{\per\KWH} grid energy rate. In future site-specific analyses, transmission line and substation capex can be included in the system cost and the grid energy rate can be refined accordingly. Mission timeframe may push costs in the other direction, as it takes time for R\&D to lower technology figures of merit to the values assumed in this work, and it is desirable to bring small missions forth sooner at greater cost. Therefore, future work should examine how small mission costs vary with technology figures of merit. It may be possible to extrapolate existing performance trends to anticipate how long it will be before each type of mission becomes affordable.

A downside to smaller cost-optimized missions is that acceleration tends to take hours to days. For missions in the ecliptic plane, beam durations of more than a few hours imply a handover from one laser to the next as the Earth rotates, whereas missions accelerating near the direction of the north or south celestial poles will never see their accelerating laser set over the horizon, and can therefore spend days to reach higher energy using a single laser. Small missions also tend to have an order of magnitude larger lightsails than the Centauri mission's \qty{4}{\meter} diameter lightsail, so future models might include a mass overhead to account for seams and joins.

System energy efficiency can be surprisingly high for relativistic missions, tending to 22\% as missions tend toward light speed. This particular limit is a function of the technology figures of merit; higher-performance materials and improved laser efficiency can further increase it. A lightsail sees its accelerating laser redshift as speed increases, which may reduce the fraction of light that it reflects. Therefore, the highest performing (and most energy efficient) highly-relativistic lightsails will need to alter their reflective properties as acceleration progresses, or the accelerating laser will need to shift bluer (perhaps in discrete steps), or both. This behavior could be factored into future models.

Looking further to the future, the performance maps also reveal a family of cost-optimal missions that accelerate at Earth gravity. The heaviest such mission is a \pDOSIIDia\ diameter \pDMass\ vessel (equivalent to 225 International Space Stations) that is laser accelerated for \pDODurPul\ to achieve \pDSpeedc, reaching the Centauri system within a human lifetime. At this scale, almost all the lightsail's mass resides in its payload and not the sail, so the sail might instead be regarded as an optical coating (and a diameter requirement) that is applied to the payload. While unthinkable at this time, the required \pDOPower\ peak radiated power (twice terrestrial insolation) might be generated by space solar power or fusion within a few centuries. The affordability of this mission's energy, raw materials, and labor, undoubtedly scale with civilization's capacity to generate energy and to automate away touch labor. Regardless of whether human civilization reaches this point or goes another way, it is now possible to contemplate such missions using laser-accelerated lightsails.

\section{Acknowledgments}
This work was supported by the Breakthrough Prize Foundation.
\iftoggle{preprint}{\bibliographystyle{elsarticle-num} }{}
\bibliography{K:/Databases/Bibtex/kp}
\end{document}